
\magnification=1200
\baselineskip=20pt
\hsize=16 truecm
\vsize=24 truecm

\font\it=cmti10

\font\title=cmbx10 scaled \magstep3

An important issue in string theory is the study of threshold
corrections in gauge couplings [1]. In particular their dependence
on the moduli of the compactified space plays a crucial role
in string unification, the derivation of the low energy effective
Lagrangian, and the problem of supersymmetry breaking. This
dependence was studied extensively in the case of symmetric orbifold
compactification of the heterotic superstring for untwisted moduli
[1],[2], and there was found that it satisfies a non renormalization
theorem [3]. Namely, it is given entirely at one-loop level and it
is determined by the violation of the integrability condition
of the corresponding $\Theta$-angles with respect to moduli. This
phenomenon
has also been understood in the field theory framework as an infrared
effect due to propagation of massless fermions in one-loop anomalous
graphs [2],[4].
A general derivation, from the point
of view of the internal $N=2$ superconformal
theory, is presented in [5].

In this note we extend the analysis of [1],[2],[3],
to the study of threshold
corrections in gravitational couplings.
In analogy to the case of
gauge couplings, to avoid infrared divergences, we consider the on-shell
three point function of two gravitons $h_{\mu\nu}$, $h_{\lambda\rho}$
and one modulus field $T$, with momenta $p_1,p_2$ and $p_3$ respectively:

$$\langle h^{\mu\nu}(p_1) h^{\lambda\rho}(p_2) T(p_3)\rangle=
\int \langle V_{h}^{\mu\nu}(p_1,z_1) V_{h}^{\lambda\rho}(p_2,z_2)
V_{T}(p_3,z_3)\rangle.\eqno(1)$$
\noindent
In (1) the integration is extended over the moduli space of
the world-sheet surface as well as over the positions of the
graviton and modulus vertex operators $V_h$ and $V_T$. There
is also an implicit summation over all even or odd spin
structures corresponding to the CP-even or CP-odd part of
the amplitude. The vertex operators for genus $g\geq 1$ read:

$$\eqalignno{V_{h}^{\mu\nu}&=:\overline{\partial}X^{\mu}(\partial X^{\nu}
+ip\cdot \psi\psi^{\nu})e^{ip\cdot X}:,&(2a)\cr
V_{T}&=v_{IJ}:\overline{\partial}X^{I}(\partial X^{J}+
ip\cdot\psi\psi^{J})e^{ip\cdot X}:,~CP~even,&(2b)\cr
V_{T}&=v_{IJ}:\overline{\partial}X^{I}\psi^{J}e^{ip\cdot X}:,
{}~g=1,~CP~odd,&(2c)\cr
v_{IJ}&=\partial_{\langle T\rangle}(G_{IJ}+B_{IJ}),&(2d)\cr}$$

\noindent
The $\psi$ 's are the left moving fermionic superpartners of the
bosonic coordinates $X$'s, and $G_{IJ}$ and $B_{IJ}$ are the
metric and the two-index antisymmetric tensor of the six-dimensional
internal space, respectively. The different form of the vertex operator
$V_{T}$ between (2b) and (2c) is due to the existence of a conformal
Killing spinor for the genus one case of odd spin structures,
which implies also a "picture changing" of the form (2b), as well as
the inclusion of one supercurrent insertion $T_{F}$  associated
to the corresponding supermodulus:

$$T_{F}=:\psi_{\mu}\partial X^{\mu}+G_{KL}\psi^{L}\partial
X^{K}:+ghosts.\eqno(3)$$

Let us first observe that the two derivative part of the amplitude
(1) vanishes for $g= 1$, because every graviton leg
contributes at least two powers of external momenta. Consider
for instance the zeroeth and first order terms in the
momentum expansion of the graviton vertex operator $V_{h}^{\mu\nu}$

$$V_{h}^{\mu\nu}={\overline{\partial}}_{z_1}X^{\mu}[\partial_
{z_{1}}X^{\nu}+ip_{1}\cdot X\partial_{z_{1}}X^{\nu}+ip_{1}
\cdot\psi\psi^{\nu}]+\it O(p_{1}^{2}).\eqno(4)$$
\noindent
Then, only the part of the amplitude where the $\psi$'s are
contracted can in principle contribute, due to the summation
over spin structures, and one is left essentially with a term
of the kind:
$${\langle\overline{\partial}}_{z_1}X(z_{1})
{\overline{\partial}}_{z_2}X(z_{2})\rangle
\rangle(\langle\psi(z_{1})\psi({z_2})\rangle)^2.$$
\noindent
Here again summation over spin structures removes
the pole of the fermionic correlator and one can safely partial
integrate, so that even this term
actually gives vanishing contribution to (1).

Therefore, there is no T-dependent, one-loop
correction to the Planck mass
and the amplitude (1) contributes only to four derivative
gravitational terms. These are quadratic in the Riemann tensor
and their structure is very similar to that of the gauge kinetic
terms\footnote{$^{1}$}{\rm Unitarity arguments [6] forbid the appearence
of other combinations of $R^{2}_{abcd}$,$R^{2}_{ab}$, and
$R^2$.}:

$$\it L_{gr}=\Delta^{gr}(T,\overline{T})R\wedge R^{\ast}+
\Theta^{gr}(T,\overline{T})R\wedge R.\eqno(5)$$
\noindent
The CP-even part $R\wedge R^{\ast}=R^{2}_{abcd}-
4R^{2}_{ab}+R^2$ is the topological Gauss-Bonnet combination
whose integral gives the Euler number, the CP-odd part $R\wedge R
=\epsilon^{abcd}R_{abef}R_{cd}^{ef}$ is the gravitational
analog of $F\wedge F$. The on-shell three-point function (1) gives
rise to $\Delta^{gr}_{T}(\langle T\rangle,\langle\overline{T}\rangle
)\equiv\partial_{\langle T\rangle}\Delta^{gr}(\langle T\rangle,
\langle\overline{T}\rangle)$ and $\Theta^{gr}_{T}$, which are
related by $N=1$ supersymmetry as follows:

$$\partial_{\langle T\rangle}\Delta^{gr}+i\Theta^{gr}_{T}=0,
{}~~\partial_{\langle\overline{T}\rangle}\Delta^{gr}-i\Theta^{gr}
_{\overline{T}}=0.\eqno(6)$$
\noindent
In fact, supersymmetry relates (1) to the three-point fermionic
amplitude of a graviton, gravitino and fermion partner of $T$,
whose chirality implies the complexification (6) of the original
bosonic amplitude (1). Futhermore if $\Theta^{gr}(\langle T
\rangle,\langle \overline{T}\rangle)$ is integrable,
i.e. it satisfies:

$$\partial_{\langle\overline{T}\rangle}\Theta^{gr}_{T}-
\partial_{\langle T\rangle}\Theta^{gr}_{\overline{T}}=0,\eqno(7)$$
\noindent
then (6) implies that $\Delta^{gr}$ and $\Theta^{gr}$ are the
real and imaginary parts respectively of an analytic function
of the moduli $T$.

The similarity of the above expressions to those of the gauge
kinetic terms allows one to extend the proof of the non
renormalization theorem of [3] to the gravitational case. This
can be done in a straightforward way in two steps: firstly all
analytic corrections must vanish because they are inconsistent
with duality invariance. Thus,the only possible corrections are
those which violate the integrability condition (7). Secondly,
the integrability condition is satisfied in all loops $g\geq 1$
pointwise in the world-sheet moduli, which shows that there are
no higher loop corrections.
\noindent
At the one-loop level the calculation can be carried out
explicitely or, alternatively, the result can be determined
through the violation of the integrability condition, using
again duality invariance. Below we present the one-loop
calculation for the gravitational couplings in (5).

Following the same arguments used above to prove that each graviton
contributes at least two powers of momenta, and using the on-shell
relations $p_i\cdot p_{j}=0$, it is easy to show that non vanishing
contributions arise only if ${\overline{\partial}}_{z_1}X^{\mu}$
is contracted with $e^{{ip_{2}}\cdot X(z_{2})}$ and ${\overline
{\partial}}_{z_2}X^{\lambda}$ with $e^{{ip_{1}}\cdot X(z_{1})}$,
yielding the correlators:

$$\langle{\overline{\partial}}_{z_1}X^{\mu}(z_{1})p_{2}\cdot
X(z_{2})\rangle\langle{\overline{\partial}}_{z_2}X^{\lambda}
(z_{2})p_{1}\cdot X(z_{1})\rangle=-p_{2}^{\mu}p_{1}^{\lambda}
\langle{\overline{\partial}}_{z_1}X(z_{1})X(z_{2})\rangle^2.
\eqno(8)$$
\noindent
Also the pure bosonic part of the graviton vertex operator
vanishes because of the summation over fermionic spin structures,
yielding the fermionic correlators:

$$\langle p_{1}\cdot\psi(z_{1})\psi^{\rho}(z_{2})\rangle\langle
\psi^{\nu}(z_{1})p_{2}\cdot\psi(z_{2})\rangle=\cases{p_{1}^{\rho}
p_{2}^{\nu}\langle\psi(z_{1})\psi(z_{2})\rangle^{2},&{\it CP~even},\cr
-\epsilon^{\nu\rho\alpha\beta}p_{1,\alpha}p_{2,\beta},&{\it CP~odd}.\cr}
\eqno(9)$$
\noindent
Finally $\overline{\partial}X^I$ in $V_T$ cannot be contracted
and thus only its zero mode part contributes; the result is then
proportional to:

$$v_{IJ}\overline{\partial}X^{I}\partial X^{J},~CP~even,\eqno(10a)$$

$$v_{IJ}G_{KL}\langle\psi^{J}\psi^{L}\rangle\overline{\partial}X^{I}
{\partial}X^{K},~CP~odd,\eqno(10b)$$
\noindent
where in (10b) the supercurrent insertion (3) was used.

{}From eqs. (10) it follows that non vanishing contribution is
possible only from sectors which preserve $N=2$ spacetime
supersymmetry. Indeed, in $N=1$ sectors all internal coordinates
$X^{I}$ are twisted by the action of the orbifold group and
consequently they do not have zero modes. On the other hand
, $N=4$ sectors give vanishing contribution in the CP even part
because of the fermionic spin-structure summation, and in the
CP odd case because of the six internal fermionic zero-modes
which cannot be saturated. In the $N=2$ sectors, only one
internal complex plane remains untwisted and $X^{I}$ have
momenta in the direction of the corresponding untwisted
torus. Taking apart the momentum dependence and performing
the two trivial $z$-integrations, one finds:

$$\Theta^{gr}_{T}={{v_{IJ}\epsilon^{JL}G_{KL}}\over{4{(2\pi)}^4
{|G|}^{1/2}}}\int{{d^{2}z d^{2}\tau}\over{{\tau_{2}}^2}}
\langle\overline{\partial}X(z)X(0)\rangle^{2}Z^{IK}_{torus}
Z_{rest}(\overline{\tau}),\eqno(11)$$
\noindent
where $\tau=\tau_{1}+i\tau_{2}$ is the modulus of the worldsheet
torus, and

$$Z^{IJ}_{torus}=\sum_{p_{L},p_{R}}p^{I}_{L}p^{K}_{R}
e^{i\pi\tau p^{2}_{L}}e^{-i\pi\overline{\tau}p^{2}_{R}}$$
\noindent
in terms
of left and right moving momenta $(p_{L},p_{R})$ of the untwisted
complex plane, and we used the relation $\langle\psi^{J}\psi^{L}
\rangle=\epsilon^{JL}$. $Z_{rest}$ is purely antiholomorphic
as a consequence of the cancellation between the contribution
of the two trasverse left-moving spacetime coordinates $X^{\mu}_{L}$
and the two internal untwisted coordinates $X^{I}_{L}$, and the
contribution
of their fermionic superpartners $\psi^{\nu}$ and $\psi^{I}$.
The last integration in (11) can be performed explicitely,
giving the result:

$$\int d^{2}z\langle\overline{\partial}X(\overline{z})X(0)\rangle
^{2}=-2\pi i\tau_{2}{d\over{d\tau_{2}}}\log(\tau_{2}\overline{\eta}
^{2}(\tau)),\eqno(12)$$
\noindent
where $\eta(\tau)=q^{1/12}\prod_{n\geq 1}(1-q^{2n})$, with $q=e^{i\pi
\tau}$, is the Dedekind ${\eta}$-function. Substituting (12) into (11) and
using the explicit form of the two dimensional lattice momenta, one
finally finds:

$$\Theta^{gr}_{T}={i\over{16 \pi^2}}\partial_{\langle T\rangle}
\int{d^2 \tau\over{\tau_{2}}}Z_{torus}{d\over{2\pi i d\overline
{\tau}}}\log(\tau_{2}\overline{\eta}^2) Z_{rest}(\overline{\tau}),
\eqno(13)$$
\noindent
where $Z_{torus}=\sum_{p_L, p_R}q^{p^{2}_{L}}\overline{q}^{p^{2}_{r}}$.

Following the arguments of ref.[1], if one considers differences
between the gravitational and gauge couplings, the $\tau_2$ dependence
inside the logarithm of (13) drops. Furthermore, the remaining
antiholomorphic function ${d\over{2\pi i d\overline{\tau}}}(\log
\overline{\eta}^2)Z_{rest}$ is modular invariant and can have at most
a singularity of the kind $\overline{q}^{-2}$ as
$\tau_{2}\rightarrow\infty$;
if this is the case it implies
that it is the $j$-invariant up to an additive constant, but in any case
$\tau_1$ integration picks the constant term, $\hat{b}^{gr}$, in its
q-expansion. $\Delta^{gr}$ can then be determined using target space
duality invariance from the violation of the integrability condition:

$$\partial_{\overline{T}}\Theta^{gr}_{T}-\partial_{T}\Theta^{gr}_
{\overline{T}}={i \hat{b}^{gr}\over{16{\pi}^2}}\int {{d^2\tau}\over
{\tau_2}}\partial_{T}\partial_{\overline{T}}Z_{torus}.\eqno(14)$$
\noindent
Using the relation $\partial_{T}\partial_{\overline{T}}Z_{torus}=
{{4\tau_{2}}\over{(T+\overline{T})^2}}\partial_{\tau}
\partial_{\overline{\tau}}(\tau_{2}Z_{torus})$, the integral (14)
becomes a boundary integral getting contributions only from massless
states. The result is:

$$\partial_{\overline{T}}\Theta^{gr}_{T}-\partial_{T}\Theta^{gr}
_{\overline{T}}={1\over{16\pi^{2}}}{{i\hat{b}^{gr}}\over{(T+\overline{T})^2}
}
,\eqno(15)$$
\noindent
which finally yields: $\Delta^{gr}={{\hat{b}^{gr}}\over{32\pi^2}}\log(T+
\overline{T})|\eta(iT)|^4$.

In the case of gauge couplings the constant $\hat{b}$ is identified
with the ${\beta}$-function of the $N=2$ theory, which , by comparing
with the CP-even part, is given by the $q^0$ term of the expression:

$$\hat{b}_i=\lim_{\tau_{2}\rightarrow\infty}Tr Q^{2}_{h}\overline{Q}
^{2}_{i}q^{H-1}\overline{q}^{\overline{H}-2}.\eqno(16)$$
\noindent
In (16) $\overline{Q}_i$ is the gauge group generator corresponding
to to the zero mode part of the Kac-Moody current and $Q_h$ is the
helicity operator emerging from the integration of $\langle\psi(z)
\psi(0)\rangle^2$ in (9). $Q^2_h$ can therefore be replaced by
${-d\over{i\pi d\tau}}\log Z_{\psi}$, where $Z_{\psi}$ is the
contribution of the two trasverse space-time fermionic coordinates.
Using that $Z_{\psi}=-{{\theta_{2}}\over{\eta}}$ or ${\theta_{3}\pm\theta_
{4}}\over{2\eta}$ for spacetime fermions\footnote{$^{2}$}{\rm The minus
sign stands for the spin-statistics.} or bosons, one finds the familiar
one loop formula:

$$\hat{b}_i={1\over{12}} Tr_{S}Q^2_i-{11\over{6}}Tr_{V}Q^2_i+
{1\over{3}}Tr_{F}Q^2_i,\eqno(17)$$
\noindent
where S,V and F stand for scalars, vectors and fermions respectively.

In the case of the gravitational coupling $Q^2_i$ is replaced
by the right moving "helicity" operator $\overline{Q}^2_h$, which
is given by ${d\over{i\pi d\tau}}\log \overline{\eta}^2$ as discussed
above. In this case one finds:

$$\hat{b}^{gr}=({1\over{12}}N^L_S-{11\over{6}}N^L_V+{1\over{3}}N^L_F)
({1\over{6}}N^R_S-{11\over{3}}N^R_V)+c(n_B-n_F),\eqno(18)$$
\noindent
where $N^L_{S,V,F}$ and $N^R_{S,V,R}$ count the corresponding left
and right moving degrees of freedom, and the last term proportional
to the difference between the total numbers of bosons and fermions
has been added, because in the derivation of the above formula
space-time supersymmetry has been used. Using now the relations
$N^L_S N^R_S=N_S$, $N^L_F N^R_S=N_F$ and $N^L_S N^R_V + N^L_V
N^R_S =N_V$ with $N_S,N_F,N_V$ the number scalars ,two-component
fermions and vector bosons respectively, and choosing $c={1\over{120}}$
one finds:

$$\hat{b}^{gr}={1\over{45}}(N_S+{7\over{4}}N_F-13N_V-
{113\over{2}}N^L_F N^R_V
+304N^L_V N^R_V).\eqno(19)$$
\noindent
Eq.(19) reproduces correctly the known trace anomaly coefficients
\footnote{$^{3}$}{\rm These are the coefficients of the square of the
Riemann tensor in the expansion of the trace of the energy-momentum
tensor.} for scalars, Majorana fermions and vector bosons. Furthermore
the coefficient of $N^L_F N^R_V$ counts for the contibution of a
gravitino together with a Majorana fermion and it agrees with the
trace anomaly of a spin-$3/2$, which is $-233/4$ in units where
a scalar contributes one. Finally, the coefficient of $N^L_VN^R_V$
counts for the contibution of the graviton, dilaton and two
index antisymmetric tensor, and it agrees with the trace anomaly
of the spin-$2$ $(212)$, of a scalar $(1)$, and of a rank-two
antisymmetric tensor $(91)$ [7].

In conclusion, we have shown that threshold corrections for gravitational
couplings, for the orbifold compactification of the heterotic string,
have the same structure and the same origin, i.e. the $N=2$ sector, as
those
for gauge couplings. Also, as a technical remark, one can notice that
the above calculation provides a relatively simple derivation, if compared
to
field theory case [8], of the trace anomaly coefficients of the various
fields
coupled to gravity.

\vskip 25pt
\noindent{\bf Acknowledgements}

I.A. and K.S.N. thank ICTP and Ecole Polytechnique respectively,
for hospitality during the completion of this work. One of us (I.A.) thanks
S.Ferrara for discussions.

\vfill\eject

\noindent{\bf REFERENCES}

\medskip

\item{[1]} V.S.Kaplunovsky, Nucl.Phys.{\bf B 307} (1988),145.
\item{[2]} L.J.Dixon, V.S.Kaplunovsky and J.Louis, Nucl.Phys.{\bf B 355}
(1991),649.
\item{[3]} I.Antoniadis, K.S.Narain and T.R.Taylor, Phys.Lett.{\bf B 267}
(1991),37.
\item{[4]} J.-P.Derendinger, S.Ferrara, C.Kounnas and F.Zwirner, {\it
On loop corrections to string effective field theories: field dependent
couplings and sigma-model anomalies}, preprint CERN-TH.6004/91(1991).
\item{[5]} I.Antoniadis, E.Gava and K.S.Narain,in preparation.
\item{[6]} B.Zwiebach, Phys.Lett.{\bf B 156} (1985),315.
\item{[7]} S.J.Gates, M.Grisaru, M.Ro\v{c}ek, W.Siegel,(1983),{\it
Superspace or One Thousand and One Lessons in Supersymmetry}, (Benjamin/
Cummins).
\item{[8]} S.M.Christensen, M.J.Duff, Nucl.Phys.{\bf B 170}[FSI]
(1980),480.

\vfill\eject

\magnification=\magstep1
\hoffset .8truecm
\voffset .7truecm
\hsize=14.5 truecm
\vsize=22.5 truecm
\baselineskip=14pt
\overfullrule=0pt
\nopagenumbers
\vskip 35pt
\rightline {IC/92/50/}
\rightline {CPTH-A160.0392}
\vskip 35pt
\centerline{\bf MODULI CORRECTIONS TO GRAVITATIONAL COUPLINGS }
\centerline{\bf FROM STRING LOOPS}
\vskip 70pt
\centerline{{\bf I.Antoniadis}$^{1}$ , {\bf E. Gava}$^{2,3}$ ,
{\bf K.S. Narain}$^{3}$}
\vskip 2.truecm
\centerline{\item 1{\it Centre de Physique Th\'eorique, Ecole
Polytechnique,
F-91128 Palaiseau, France}}
\centerline{\item 2{\it Istituto Nazionale di Fisica Nucleare-sez.
 di Trieste, Italy}}
\centerline{\item 3{\it International Centre for
 Theoretical Physics, I-34136 Trieste, Italy}}
\vskip 90pt
\centerline{\bf Abstract}
\noindent
We study moduli dependent threshold corrections
to gravitational couplings in the case
of the heterotic string compactified on a symmetric orbifold,
for untwisted moduli, extending previous analysis on gauge couplings.
Like in the gauge case, the contribution comes entirely from the
spacetime $N=2$ sector. As a byproduct, this calculation provides
a simple derivation of the trace anomaly coefficients for the
different fields coupled to gravity.

\vfill
\centerline{February 1992}
\vfill
\eject
\bye